\documentclass[acmsmall]{acmart}
\usepackage{enumitem}
\AtBeginDocument{%
  \providecommand\BibTeX{{%
    \normalfont B\kern-0.5em{\scshape i\kern-0.25em b}\kern-0.8em\TeX}}}

\setcopyright{rightsretained}
\acmJournal{PACMHCI}
\acmYear{2022} \acmVolume{6} \acmNumber{CSCW2} \acmArticle{378} \acmMonth{11} \acmPrice{}\acmDOI{10.1145/3555103}

\usepackage{xspace}
\usepackage{csquotes}
\usepackage{caption}
\usepackage{subcaption}
\definecolor{darkcyan}{rgb}{0.0, 0.55, 0.55}

\newcommand{\salim}[1]{{#1}}

\newcommand{\tadv}{\textit{timeline-advertisers}\xspace}
\newcommand{\padv}{\textit{adsettings-advertisers}\xspace}
\newcommand{\uadv}{\textit{unmatched-advertisers}\xspace}
\newcommand{\madv}{\textit{matched-advertisers}\xspace}

\begin{document}

%%
%% The "title" command has an optional parameter,
%% allowing the author to define a "short title" to be used in page headers.
%\title{An Analysis of Micro-Targeting Practices by Small and Large Companies on Facebook}
\title{Exploring the Online Micro-targeting Practices of Small, Medium, and Large Businesses}

%Who dominates micro-targeted online advertising? An Analysis of Micro-Targeting Practices by Small and Large Companies on Facebook}

%%
%% The "author" command and its associated commands are used to define
%% the authors and their affiliations.
%% Of note is the shared affiliation of the first two authors, and the
%% "authornote" and "authornotemark" commands
%% used to denote shared contribution to the research.

\author{Salim Chouaki}
\affiliation{
  \institution{Univ. Grenoble Alpes, CNRS, Grenoble INP, LIG}
  \country{France}}
\email{salim.chouaki@univ-grenoble-alpes.fr}

\author{Islem Bouzenia}
\affiliation{
  \institution{Univ. Grenoble Alpes, CNRS, Grenoble INP, LIG}
  \country{France}}
\email{islem.bouzenia@univ-grenoble-alpes.fr}

\author{Oana Goga}
\affiliation{
  \institution{Univ. Grenoble Alpes, CNRS, Grenoble INP, LIG}
  \country{France}}
\email{oana.goga@univ-grenoble-alpes.fr}

\author{Béatrice Roussillon}
\affiliation{
  \institution{Univ. Grenoble Alpes, GAEL}
  \country{France}}
\email{beatrice.Roussillon@univ-grenoble-alpes.fr}

%%
%% By default, the full list of authors will be used in the page
%% headers. Often, this list is too long, and will overlap
%% other information printed in the page headers. This command allows
%% the author to define a more concise list
%% of authors' names for this purpose.
%\renewcommand{\shortauthors}{Islem et al.}

\renewcommand{\shortauthors}{Salim Chouaki et al.} %% No italics %%

%%
%% The abstract is a short summary of the work to be presented in the
%% article.

\begin{abstract}
Facebook and other advertising platforms exploit users' data for marketing purposes by allowing advertisers to select specific users and target them with well-crafted messages (the practice is being called micro-targeting). However, advertisers such as Cambridge Analytica have maliciously used these targeting features to manipulate users in the context of elections. The European Commission plans to restrict or even ban some targeting functionalities in the new European Democracy Action Plan act to protect users from such harms. The difficulty in finding appropriate restrictions is that we do not know the economic impact of these restrictions on regular advertisers.
In this paper, to inform the debate, we take a first step by understanding who is advertising on Facebook and how they use the targeting functionalities of ad platforms. For this, we asked 890 U.S. users to install a monitoring tool on their browsers to collect the ads they receive on Facebook and information about how these ads were targeted. By matching advertisers on Facebook with their LinkedIn profiles, we could see that 71\% of advertisers are small and medium-sized businesses with 200 employees or less, and they are responsible for 61\% of ads and 57\% of ad impressions.
Regarding micro-targeting, we found that only 32\% of small and medium-sized businesses and 30\% of large-sized businesses in our dataset micro-target at least one of their ads. 
These results should not be interpreted as micro-targeting not being useful as a marketing strategy, but rather that advertisers prefer to outsource the micro-targeting task to ad platforms. Indeed, to deliver ads, Facebook is employing optimization algorithms that exploit user data to decide which users should see what ads; which basically means ad platforms are performing an algorithmic-driven micro-targeting. 
Hence, when setting restrictions on micro-targeting, legislators should take into account both the traditional advertiser-driven micro-targeting as well as algorithmic-driven micro-targeting performed by ad platforms.

\end{abstract}


%%
%% Keywords. The author(s) should pick words that accurately describe
%% the work being presented. Separate the keywords with commas.

%%
%% This command processes the author and affiliation and title
%% information and builds the first part of the formatted document.
\maketitle

%!TEX root = main.tex

\section{Introduction}

Digital advertising is an industry of billions of dollars that is outperforming other traditional advertising industries. According to Statista, 356 billion U.S. dollars were spent on digital advertising in 2020~\cite{Statista2021} while only 172 billion dollars were spent on offline advertising~\cite{statista_offline_data}. 
The enormous success of online advertising platforms is generally justified by the precise targeting features they offer~ \cite{facebook_facebook_2021}.  
Ad platforms exploit the data they collect on users to infer their \textit{interests} and \textit{demographics}, allowing advertisers to target \textit{audiences} with precise characteristics. For instance, advertisers can target groups of users as small as hundreds and as specific as ``people interested in <<Environmental impact of meat production>> that have a <<Doctoral degree>> and who recently moved to Paris''. Facebook proposes to advertisers over 240k such attributes to select their audiences~\cite{speicher_potential_2018}. 
We call this form of targeting, \textbf{advertiser-driven micro-targeting} as the advertisers themselves select the precise characteristics of the audience they want to reach. A complementary and lesser-known type of micro-targeting is \textbf{algorithmic-driven micro-targeting}, where ad platforms themselves exploit user data (going beyond user's interests and demographics) to decide which users should receive what ads. For example, when delivering ads, Facebook computes how ``relevant'' ads are to specific users and which users will bring more value if they see the ad~\cite{adrelevance,Ali2021,Ali2019a}. Effectively, it is not the advertiser but the ad platform that decides the right audience for a particular ad. Algorithmic-driven micro-targeting can be done standalone or in conjunction with advertiser-driven micro-targeting~\cite{facebook_business_about_ad_delivery}.

From an \textit{economics perspective}, improving the matching between consumers and products through micro-targeting increases consumer surplus and decreases producer costs~\cite{goldfarb2014different, bergemann2011targeting}. Consumers save time and effort searching for the right product and information, while 
producers, obtain a high return on investment by reaching interested consumers at small costs. In addition, micro-targeting is seen as a marketing tool that levels up the competition between large-seized businesses (BEs) and small and medium-sized businesses (SMEs), leading to a more fair market~\cite{goldfarb2014different,goldfarb2019digital}.

On the other hand, form a \textit{security and privacy perspective}, micro-targeting can be weaponized by malicious third parties to influence user's decisions, by targeting messages that resonate well with users beliefs and personalities~\cite{qz-oldconservatives,wong2019cambridge,shane2017these,Lewandowsky2020,Susser2019}. 
The Cambridge Analytica scandal and the Russian interference in the U.S. 2016 election are concrete examples of how micro-targeting can be used to engineer polarization, promote voter disengagement, or manipulate voters~\cite{shane2017these,IRA2,wong2019cambridge,confessore2018cambridge}.

As a result, many governments and regulators are studying how to regulate online advertising and micro-targeting. For example, the European Commission is currently working on two acts: the Digital Services Act and the European Democracy Action Plan~\cite{DSA, EDAP}. One controversial solution on the European Commission's table is to \textbf{ban altogether or restrict the ability to micro-target ads online}, especially if they contain political messages~\cite{jourova}. Another solution supported by many privacy advocacy groups is to restricting web tracking, limiting this way the possibility to build user profiles.\footnote{Solution supported by the European Digital Rights (EDRi) in their report ``Targeted online: An industry broken by design and by default'' and governmental agencies such as the Norwegian Consumer Council in their report ``Time to ban surveillance-based advertising''~\cite{edri, ncc}.}
While these restrictions are desirable from a security and privacy perspective, \textbf{the difficulty in finding appropriate regulations is that we do not know their economic impact on regular advertisers} and how they will impact their economic growth, customer acquisition, and, more broadly, the fairness of the market.
For example, banning micro-targeting might unevenly impact small and medium-sized businesses compared to large-sized businesses, and Facebook has brought harming small-sized businesses as the main argument against the new Apple's iOS 14 restrictions on cookie tracking~\cite{techcrunch1,fb_small_businesses}. 
While these economic considerations are a barrier in adopting regulations, there is a lack of data and studies on the use of micro-targeting technologies by small, medium, and large-sized businesses and the usefulness of web tracking.

In this paper, we take a first step toward understanding the potential economical impact of such restrictions by \textbf{exploring to which extent small and medium-sized businesses make use of online advertising and analyzing which micro-targeting practices they employ}. 
We focus this study on Facebook as it is one of the most used online advertising platform and accounts for 19.3\% of the market share~\cite{times_facebook_nodate}. 
To audit advertisers and their micro-targeting, we developed a monitoring tool that users can install on their browsers. This monitoring tool observes and sends, to our servers, the ads seen by users in their Facebook timeline and the targeting options used by the advertisers to target the ads sourced from the ``Why am I seeing this ad?'' feature~\cite{Andreou2018} (Section~\ref{sec:data}). 
Using a crowdsourcing platform, we recruited 890 users across multiple states in the U.S. to install our monitoring tool for at least six weeks. In total, we collected 102k ads, coming from 40k advertisers targeted during Nov 2020 to Jan 2021. The data collection is covered by an IRB and is compliant with GDPR procedures. 
The data collected from the monitoring tool provides us with information about the name, website, and sometimes the address of the advertisers; however, it does not provide data about the size of the corresponding business. To get such information, we propose a method to match the Facebook profiles of advertisers with their corresponding LinkedIn profiles (Section~\ref{sec:data}). We estimate the accuracy of our matching method to 86.5\%. As a result, we found a LinkedIn profile for 66\% of advertisers in our database.

Our results, in Section~\ref{sec:adv}, show that 71\% of advertisers in our matched dataset (i.e., advertisers for which we found a LinkedIn profile) are SMEs with 200 employees or less, and they are responsible for 61\% of ads and 57\% of ad impressions. In addition, we observed that out of a sample  of 100 unmatched advertisers we manually checked (with no LinkedIn profile), the majority correspond to \textit{individual entrepreneurs}: 28\% corresponded to personal Facebook pages (e.g., photographer, model), 24\% were Facebook pages for individual services (e.g., house keeper, pet trainer), and another 24\% were online stores run by individuals. 
Hence, our results show that \textit{the large majority of advertisers on Facebook are small and medium-sized businesses}. This supports economic models that predict that 
the relatively low cost of micro-targeting enables smaller businesses to access advertising markets from which they were previously excluded~\cite{taiminen2015usage, bagwell2007economic, anderson2006long, goldfarb2014different}.

The second question we address in this paper is to which extent businesses employ advertiser-driven micro-targeting strategies (Section~\ref{sec:targeting}). We consider that if advertisers only specify the age, gender, and location of the users they want to reach, they are not micro-targeting their ads. Otherwise, we consider that advertisers are micro-targeting their ads if they target their ads to users that have specific attributes (e.g., users interested in organic food, users who are parents) or to users that have previously interacted with their business (e.g., users who bought from their shop, users who visited their website). Our results show that only 32\% of small and medium-sized businesses and 30\% of large-sized businesses use micro-targeting in at least one of their ads. In addition, 46\% of advertisers with no LinkedIn profile used advertiser-driven micro-targeting. Hence, these results show that \textit{most businesses do not seem to rely on advertiser-driven micro-targeting}. This is a significant shift from what we observed in our previous study in 2019 where we found that 64\% of the ads collected were micro-targeted by advertisers~\cite{andreou_measuring_2019}. 
It is important to understand, however, that these results do not imply that micro-targeting is not useful as a marketing strategy; they rather suggest that the advertisers prefer to outsource the micro-targeting task to the ad platform. 
Indeed, Facebook claims to perform algorithmic-driven micro-targeting for every ad impression~\cite{facebook_business_about_ad_delivery}. The shift toward algorithmic-driven micro-targeting could be explained by ad platforms' algorithms getting better than advertisers at finding the right audiences to target ads. This points to the increased power of targeting algorithms implemented by platforms, and brings new challenges in regulating micro-targeting, as it is mainly done implicitly by ad platforms.

We finally investigate the link between web tracking and micro-targeting on Facebook as the platform already has a lot of information about the behavior of users inside the platform (Section~\ref{sec:tracking}). 
Facebook claims that user's interests are only inferred from user's activity inside Facebook~\cite{facebook_business_about_detailed_targeting}, while ``ad relevancy'' scores are based on both inside and outside Facebook data~\cite{facebook_business_ad_principles}.
Hence, most advertiser-driven micro-targeting options (except retargeting) do not seem to rely on web tracking, while algorithmic-driven micro-targeting relies on web tracking. 
This could be a possible explanation for the popularity of algorithmic-driven micro-targeting among advertisers. 
To understand the extent to which advertisers contribute to Facebook's data acquisition through web tracking, we investigate what fraction of advertisers have installed the Facebook pixel on their website. The Facebook pixel is a piece of code that allows Facebook to know which Facebook users have visited a particular website~\cite{pixel_fb}. Our results show that 81\% of SMEs and 69\% of BEs we checked have deployed one or more Facebook pixels on their websites, hence being a widespread practice. Nevertheless, only 2.6\% of the ads sent by SMEs and 2.7\% of the ones sent by BEs were targeted explicitly with the Facebook pixel, to reach people that visited the advertiser's website. This raises questions of whether there are any benefits for advertisers from installing a Facebook pixel, besides retargeting visitors.  

These findings have several implications.  First, from a \textit{restrictions} perspective, the focus has been on restricting advertiser-driven micro-targeting; however, there needs to be a discussion on whether the algorithmic-driven micro-targeting done by ad platforms themselves is a practice that is more acceptable or should be restricted as well. Second, from a \textit{fairness} perspective, previous works have shown that algorithmic-driven micro-targeting can lead to discrimination across gender and political affiliation~\cite{Ali2021}.  Especially if advertiser-driven micro-targeting is restricted, it is important to study whether algorithmic-driven micro-targeting does not lead to unfairness across advertisers. 
Third, from a \textit{transparency} point of view, the information in the ``Why am I seeing this ad?'' features is incomplete as it only provides information about the advertiser-driven micro-targeting but not about the algorithmic-driven micro-targeting, i.e., we know the characteristics of the users the advertisers chose to target, but we do not know the characteristics of users the ad platform choses to deliver the ads to.  Given the shift towards algorithmic-driven micro-targeting, this is an important information to have. 
We hope these results will contribute to the debate on finding adapted restrictions for online advertising and steer research in devising methods to evaluate the economic impact of advertising regulations. To support future research, the list of advertisers as well as the details we collected about their businesses can be found \href{https://github.com/Salim-Chouaki/Micro-targeting-practices-on-Facebook}{on this link}\footnote{\href{https://github.com/Salim-Chouaki/Micro-targeting-practices-on-Facebook}{https://github.com/Salim-Chouaki/Micro-targeting-practices-on-Facebook}}.

%!TEX root = main.tex

\section{Background}
\label{sec:background}

\subsection{Advertising on Facebook}
At a high level, advertising on Facebook works as follows: Facebook infers the characteristics of its users; these characteristics are made available to advertisers to target their ads; upon the creation of an ad, Facebook delivers the ad to users in a way that maximizes the total value which is a combination of the advertisers profit and the user's quality of experience on the platform. Next, we describe the important notions to know about the platform's actors (e.g., users, advertisers) and the process performed or triggered by them (e.g., ad creation, ad delivery). We use quotes from Facebook's help pages, to illustrate precisely the current descriptions. 
\paragraph{Users.}
From an advertising process perspective, a Facebook user is characterized by a set of interests (e.g., pancakes, swimming, guitar) and demographics (e.g., age, gender, location). Facebook infers these characteristics based on data about the activity of the user~\cite{facebook_business_about_detailed_targeting}: 

\begin{displayquote}
\small{\em ``Ads they click;
Pages they engage with;
Activities people engage in across Meta technologies related to things like their device usage, and travel preferences;
Demographics like age, gender and location;
The mobile device they use and the speed of their network connection.''}
\end{displayquote}
Hence, while Facebook is able to observe the activity of users both inside and outside Facebook (through web tracking~\cite{pixel_fb}), Facebook claims to only use inside data to infer the interests and demographics of users~\cite{Andreou2018}.

\paragraph{Advertisers.}
To run ads on Facebook, one needs to create an advertising account.  
There are two types of advertising accounts: \textit{personal} and \textit{business}. To create a personal advertising account, the user only needs a Facebook account and a valid credit card.  A personal advertising account allows the user to create multiple Facebook pages and run ads associated with them. The associated Facebook pages do not need to be registered businesses. They can represent anything, be it commercial or non-commercial, and range from personal cooking blog and TV Show to a retail or clothing company.  
To create a business advertising account, the user needs to prove the existence of their business to Facebook, by giving the official registered number in the local business license for example. A business account can be used by a group of people and can associate multiple advertising accounts and pages. Both types of accounts can use the same targeting options, the only difference is for custom audience-based micro-targeting where a business account can have up to 100 pixels (trackers) while a personal account can only have one~\cite{facebook_business_pixel}. Furthermore, a business account offers advantages in terms of advertising management by allowing multiple people and external collaborators or contractors to create ads on behalf of the business.

When a user receives an ad on Facebook, s/he will know that the ad comes from (one of the) Facebook pages associated with the advertiser, but s/he will not know whether the advertiser behind uses a personal or business account, and what is the name of the person or business that is responsible for the ad.

\paragraph{Ad creation.}
Facebook provides advertisers with an interface to manage and create ads, called Ads Manager~\cite{ads_manager_fb}. The \emph{first} step in configuring an advertising campaign is to choose the objective of the campaign. Facebook proposes three main categories of \textit{objectives}: awareness, consideration, and conversion~\cite{objectives_fb}. Brand awareness is about reaching new people and informing them about the services of the business. Consideration campaigns target people who are most likely to consider checking the business and look for more information about it. 
Conversion campaigns target users who are most likely to buy a product or subscribe to a service. In the \emph{second} step, the advertisers need to configure the campaign in terms of \textit{budget}, \textit{running time} (start and end date), \textit{targeted audience} (the characteristics of users they want to reach), \textit{placement} of the ad (e.g., Facebook, Instagram, mobile, computer), \textit{frequency of ad delivery} (how many times a user should see the ad per day), and \textit{bid control} (the maximum budget to pay for one ad impression). \emph{Finally}, the advertiser needs to input the \emph{content} of the ad in terms of the ad text, media (e.g., image, video), landing URL (e.g., website, Facebook page), and call-for-action button (e.g., Buy now, Learn more).

\paragraph{Ad delivery.}
The Facebook ad delivery system uses \textit{ad auctions} and \textit{machine learning algorithms} to determine where, when and to whom to show an ad. These processes work together to maximize value for both users and advertisers by taking into consideration three factors:
   \setlist{nolistsep}
\begin{enumerate}
  \item \textit{Bid:} What the advertiser can pay to achieve the desired outcome.
  \item \textit{Estimated action rates:} An estimate of whether a particular person engages with or converts from a particular ad.
  \item \textit{Ad quality:} A measure of the quality of an ad as determined from many sources, including feedback from people viewing or hiding the ad and assessments of low-quality attributes in the ad, such as withholding information, sensationalized language, and engagement bait.
\end{enumerate}
   \setlist{listsep}
   \setlist{}
Facebook's estimation of how \textit{relevant} an ad is to a user plays a significant role in the actual ad delivery. In Facebook's words~\cite{facebook_business_about_ad_delivery}: 

\begin{displayquote}
\small{{\em ``Each time there's an opportunity to show an ad to someone, an ad auction takes place to determine which ad to show.'' ...``Since we want each person to see relevant ads, the ad auction considers predictions of each ad's relevance for the particular person. This means that ads with higher relevance can win ad auctions at lower costs.'' ... 
``Relevance predictions estimate how likely the person is to consider the ad high quality and take the advertiser's desired action.''}}
\end{displayquote}
An ad that is relevant to a person could win an auction against ads with higher bids. Each time an ad is displayed, the delivery system learns more about delivery optimization. Thus, the more an ad is displayed, the better the delivery system optimizes to whom to send the ad. This may have important implications regarding the fairness of competition and it is further discussed in the economics of advertising section.

\subsection{Types of micro-targeting}
In this paper, we point out that there are two broad types of micro-targeting based on whether micro-targeting is driven by the advertiser or by the ad platform's algorithms.

\vspace{2mm}
\noindent \textbf{Advertiser-driven micro-targeting}
results from the ad creation process when advertisers choose the characteristics of the audience they want to reach. 
We group the targeting options offered to advertisers in four categories: 
\begin{enumerate}
\item \textbf{Generic targeting only:} Corresponds to targeting using generic parameters such as: \textit{age}, \textit{gender}, \textit{location}, and \textit{language}. These parameters can be used in combination with other targeting parameters (see below); in this category we include ads with generic targeting only.  We do not consider this targeting as micro-targeting since these categories are very broad. 
\item \textbf{Attribute-based micro-targeting}: Advertisers can send their ads to users who match with specific attributes such as \textit{interests} (e.g., online shopping, movies, music),  \textit{behaviors} (e.g., engagement in political content, frequent travelers), or \textit{demographics} (e.g., relationship status, work, education). We define an ad as targeted using attribute-based micro-targeting when the advertiser chose one or more combinations of these attributes to target its ads.
\item \textbf{Custom-audience micro-targeting}: Advertisers can target users that previously interacted with their business through custom-audiences. An advertiser can form custom-audiences from different data sources: data files containing a list of emails or users ids, website visits detected by a hidden Facebook pixel, users who engaged with the advertiser's page, video, or Instagram profile. Re-targeting is part of this category as it contains users that already interacted with the business. We consider this as micro-targeting since the advertisers give a specific list of users, they want to reach, as input to Facebook. 
 
\item \textbf{Lookalike audience}:  While targeting existing customers is an important requirement in online advertising, expending the customer base is equally important. Facebook proposes a targeting mechanism called ``lookalike audiences'' that allow advertisers to target users similar to the ones existing in their custom audiences~\cite{facebook_facebook_2021}. When creating a lookalike audience, it is Facebook's algorithms that decide who are the similar users. Thus, the advertisers do not know what are the characteristics used by Facebook to extend the initial custom audience. 
We consider this as a form of \textit{\textbf{explicit}} \textbf{algorithmic-driven micro-targeting}, as the advertiser outsource the work of finding relevant customers to Facebook, and has little control over the characteristics of users in the lookalike audience. 
\end{enumerate}

\vspace{2mm}
\noindent \textbf{Algorithmic-driven micro-targeting}
results from the optimization processes used by the ad platform during ad delivery.  When there is an opportunity to show an ad, Facebook looks at all the ads that it could deliver to the particular user, and chooses to show the one for which it has calculated the highest ad relevancy score~\cite{score_relev}. Compared to lookalike audience, this is an \textbf{\textit{implicit}} form of algorithmic-driven micro-targeting.
Facebook markets algorithmic-driven micro-targeting as a tool to find the right customers~\cite{facebook_business_deliver_relevance_and_growth_with_facebook}:
\begin{displayquote}
\small{
{\em
``And unlike search and retargeting, we can identify a wide range of explicit and implicit intent signals to enable marketers to generate new demand and fulfill existing demand so you can reach potential valuable customers at the start of their journey...'' 

``Today's fastest growing brands don't wait for customers to come to them. They don't wait for purchase intent, they create purchase intent—and Facebook makes it easier to accurately and predictably pair the right products to the right people to generate demand and sales—at an enormous scale.''
}
}
\end{displayquote}
It is important to note that Facebook performs algorithmic-driven micro-targeting for every ad regardless of whether it has been micro-targeted, or the advertiser only used generic targeting ~\cite{facebook_business_ad_principles}:
\begin{displayquote}
\small{
{\em
``For example, imagine that the owner of a hair salon wants to reach women aged 20-60 in the city where the salon is located. She creates an ad on Facebook to show to the people in that area who are the most likely to be interested in the salon.
Our system can show the hair salon's ads to the people in that area who are most likely to be interested in the salon. Interests are determined based on activity on Facebook, such as Pages or posts that people like, posts or comments they make and activity off Facebook from apps they use and websites they visit. If someone sees the ad, it's because they're in the group of people that seems to fit this criteria set by the advertiser. This process can happen without Facebook having to tell the advertiser a person's name or contact info.''
}
}
\end{displayquote}

Facebook chooses to whom to send an ad based on the objectives of the campaign (e.g., reach, calls, store traffic, etc.), and the quality of users' experience with the advertiser's ads. 
\begin{displayquote}
\small{
{\em ``When you make your Optimization for Ad Delivery choice for an ad set, you're telling us to get you that result as efficiently as possible. In other words, your optimization choice is the desired outcome that our system bids on in the ad auction. For example, if you optimize for link clicks, your ads are targeted to people in your audience who are most likely to click the ads' links.''}
}
\end{displayquote}

Besides some high-level explanations of the importance of algorithmic-driven micro-targeting for advertisers, there is not much transparency around how precisely it is done and what are all the data sources that enter the equation (e.g., the role of web tracking or the content of the ad).

%!TEX root = main.tex
\section{Data Collection}
\label{sec:data}

Ideally, we would like analyze a representative sample of ads that users see on Facebook and how these ads have been targeted. Facebook provides an Ad Library~\cite{facebook-ad-library} where one can search whether a specific advertiser has any active ad campaigns. Sadly, the Ad Library does not provide historic data, and we can only query ads for advertisers who's names we already know; hence making it impossible to get a representative sample of ads. The Ad Library also lack information about how the ad was targeted. 
We opted to collect data through a browser extension that can observe the ads users receive on Facebook, and augment the data with information from LinkedIn.

\subsection{Data from Facebook}

We developed a browser extension\footnote{\href{https://github.com/Salim-Chouaki/CheckMyNews}{https://github.com/Salim-Chouaki/CheckMyNews}} that is available on the Chrome store and that users can install. The tool extracts all the ads a user receives on their Facebook timeline. Our tool collects the text of the ad, the media (e.g., image, video), and the advertiser who sent the ad (i.e., its corresponding Facebook page). Our tool also collects the targeting information shown in the ``Why am I seeing this ad?'' explanation. Facebook presents in this ad explanation the characteristics of the audience chosen by the advertiser~\cite{adexplanations}. It includes age, gender, location, targeted interests, and type of targeted audience (e.g., custom audience, lookalike audience); see Figure~\ref{fig:explanation} in Appendix. 
As in our previous work~\cite{andreou_measuring_2019}, to identify the ads, we search for objects that contain the pattern ``Sponsored'' in the HTML of the Facebook timeline of a user. To get the information in the ``Why am I seeing this ad?'' feature, we send a request to the Facebook API for each of the collected ads.\footnote{There are, in fact, numerous technical difficulties that come when trying to collect such data. For example, Facebook often takes an adversarial approach where they change the HTML markup of their page to break extensions such as ours. We developed a suite of strategies to cope with such changes and quickly detect if the extension no longer collects data well. We omit these details from the paper, but the code of our collection method can be found \href{https://github.com/Salim-Chouaki/CheckMyNews}{here}.}

\paragraph{Quality of ``Why am I seeing this ad?'' explanations.} In Previous work, we found that ad explanations provided by the ``Why am I seeing this ad?'' feature show only one (out of potentially many) attributes selected by advertisers to target their ads~\cite{andreou_measuring_2019}. This limitation does not affect our targeting types analysis in Section~\ref{sec:targeting} since we just need to detect one attribute to consider that the ad was micro-targeted.
Nevertheless, we launched several campaigns (reproducing the approach by Andreou et al.~\cite{Andreou2018}) where we targeted the users that installed our extension and had specific interests and demographics and investigated if the ad explanations accurately portrayed the parameters we chose to target our ads. The ad explanations correctly reflected our targeting parameters. Indeed, Facebook announced they revamped this feature, and they seem to have solved some of the limitations~\cite{facebook_news_ads_explanations}.

\paragraph{User recruitment.} We recruited 890 U.S. users on Prolific and asked them to install our extension on the computer where they usually connect to Facebook and keep it active, at least, for six weeks. \salim{To incentivize the users to follow our requests, we gave them an initial payment of 0.5£ right after the installation and a more significant bonus of 4.1£ at the end of the six weeks if they had satisfied the minimum required activity level over the six weeks}.\footnote{Payments on Prolific are in £ irrespective of the country of origin of users.} Unfortunately, many users were concerned about installing software on their computers and chose not to participate in our study. Being able to collect data from 890 users is a remarkable effort. Some users kept the extension for less than six weeks, while others kept it longer. 
Our dataset includes data we collected over three months, from Nov 2020 until the end of Jan 2021.

\paragraph{Representativeness.} While we cannot guarantee representativeness, we tried to limit biases by recruiting users from various U.S. states and ethnicities. To validate that the majority of users are from the U.S., we examined information in the ``Why am I seeing this ad?''; this ad explanation contains information about the location that the advertiser targeted. The location was set to "The United States" or a U.S. state in 99\% of the ads. We found 41 users that received at least one ad where the location was set to another country. For 16 of them, more than 70\% of their ads were not targeted to the U.S.; we ignore these users and their ads in the analysis (they represent less than 0.02\% of all ads). For the 25 other users, less than 10\% of the ads they received were targeted outside the U.S.; we consider only the ads targeted to U.S. locations. Figure~\ref{fig: usmap} (in Appendix.) presents the location of the users in our dataset mapped from their IP addresses. Moreover, we infer the age and gender of users from ad explanations. Our users are 65\% men and 35\% women (compared to 55\% men, 45\% for U.S. users on Facebook), and 70\% are between the age of 20 and 40. Figure~\ref{fig: agesdist} (in Appendix) shows the distribution of our users by age compared to the one of Facebook. 
\paragraph{High-level data description}
In total, our users received 102k unique ads from 40k different advertisers in their Facebook timelines. We call them the \textit{\tadv}. The 102k ads got 185k impressions. Impressions represent the number of times an ad has been delivered to users. For example, if a user received the same ad twice or two different users received the same ad, we say that the ad had two impressions. We differentiate between ads using the unique ID attributed by Facebook to each sent ad that we can find in the HTML content. 

We were able to collect ad explanations for 45\% of ads. The task to collect the explanation of all ads is not easy because Facebook considers it an abnormal behavior when bursting the platform with ad explanation queries, and such behavior triggers a security check from Facebook. To avoid such an annoying experience, we built a scheduler that waits five minutes between two requests of ad explanations. This leads to loose 55\% of the ad explanations.

In addition, our browser extension collects, once a day, the information present in the Ad Settings page of the user~\cite{facebook_preferences_advertisers}. This page contains information about the ``Advertisers You've Seen Most Recently''. We call them the \textit{\padv}. The \padv nicely complement our \tadv because they contain advertisers who target the user on other devices where our extension is not installed. 
There are 53k unique \padv. The overlap between \tadv and \padv is 14k, and the total number of advertisers in our dataset is 79k. Having two sets of advertisers, we obtained using two different methods, is helpful to assess the generality of our results.

\subsection{Data from LinkedIn}
To identify the size of the business of each advertiser in our dataset, we developed a method to match advertisers on Facebook with their corresponding LinkedIn profiles. LinkedIn provides information about the number of employees for each business in the about page.
 Our matching method consists of taking the list of advertisers from our dataset and searching the company name on Google by doing a query of the form: ``NAME company LinkedIn'' where NAME is the name of the advertiser we have in our database. We take the first three search results on Google; then, we look for the first search result that matches a company profile on LinkedIn by checking if the URL is in the following form: www.linkedin.com/company/name-of-company/about. 
When we find a match, we extract the company's profile page present in the \textit{about} section of the company. We found a match for 66\% of our 79k advertisers (82\% for \tadv and 50\% for \padv), which we call the \madv. We call the \uadv the remaining 34\%. 

To investigate why we did not find matches for all advertisers, we took a random sample of 100 \uadv and manually investigated if they have a LinkedIn profile. We could not find a LinkedIn profile for any of them, and they seem to correspond to personal service providers (e.g., pet training, hair cutting) that do not seem to be registered businesses.  
This means that or matching method does not only have a recall of 66\%, but that \uadv simply do not have LinkedIn profiles.

To investigate the precision of our matching method, we randomly selected 200 Facebook advertisers for which our method found a match, and we manually verified if the match was correct. The match was correct in 86.5\% of the cases. The mismatches correspond mostly to advertisers without a LinkedIn profile for which our algorithm matched with the company that has the most similar name. 
Furthermore, we investigated the matching accuracy across businesses of different sizes by taking separate samples of 100 ads for small, medium, and large-sized businesses. 
We found out that most mismatches happen with small-sized businesses (25\% error rate) while medium and large-sized businesses have lower error rates (4\% for medium-sized businesses, 1\% for large ones). The majority of small-sized businesses wrongly matched, are matched with other small-sized businesses.

\subsection{Data collection compliance and limitations}

\paragraph{IRB and Ethic considerations} We only collected information about the ads and the corresponding ad explanations. We did not collect any personally identifiable information such as names, emails, or phone numbers of users, and we did not collect their personal messages or personal posts. We clearly stated what data we collected to the Prolific users, and before they installed the extension, they had to sign a consent. 
All data collection that we present in this paper was reviewed and approved by the Ethical Review Board of Inria and approved by the local Data Protection Officer. The data collection is GDPR compliant, and users had to give explicit consent (that has been previously validated by the Data Protection Officer) to participate in the study, and they were able to leave and ask for their data to be removed at any point in the study. Our data collection does violate the Terms of Service of Facebook. Our IRB committee is aware of this and has decided that the benefits of the research we conduct are sufficiently justified to collect such data. 
The code of our browser extension is open source as it can be viewed in the client's machine, as any Chrome and Firefox extension. Data collections such as ours were violating Prolific's ToS in October 2020. 
We contacted Prolific, and they agreed to allow us to run the experiment under their strict supervision. Prolific is now allowing other researchers to perform such tasks~\cite{prolific_download_software}. 

\paragraph{Limitations}
Even though the data collection we propose in our work is extremely valuable and provides a unique opportunity towards analyzing advertisers and their practices on Facebook. It also has some limitations:
\begin{trivlist}
\item \hspace{3mm} (1) Our work is mainly based on data collected from a desktop browser extension that can capture the ads users receive on Facebook. However, many users also use Facebook through their phones and receive ads that we cannot collect. 
It is technically challenging (if not impossible) to provide such a tool for mobile phones. However, to provide a complete picture, we also analyze advertisers that appear in the ad settings pages of users (i.e., the ``Advertisers you've seen most recently''), which contains the names of advertisers that targeted the users on other devices. 
\item \hspace{3mm} (2) We had to use a scheduler that waits five minutes between two ad explanation requests to avoid security checks from Facebook because bursting the platform with ad explanation queries is considered abnormal behavior. As a result, we could only collect the ad explanation for 45\% of ads.   
The missing explanations are randomly spread across users, advertisers, and time. We did not observe any biases between ads for which we have explanations and ads for which we do not. 
\item \hspace{3mm} (3) Our study is limited to U.S. users and mostly U.S. advertisers. However, we make our code for collecting ads publicly available which might help further research in other countries and populations.  
\item \hspace{3mm} (4) Ad explanations from the ``Why am I seeing this ad?'' might have limitations. However, we described that previously known limitations do not affect our analysis. In addition, we performed several controlled experiments, and the ad explanations correctly presented our targeting parameters. 
\item \hspace{3mm} (5) While we cannot guarantee the representativeness of users and advertisers, we tried to limit biases by recruiting users from various U.S. states, genders, and ethnicities.   
\end{trivlist}

\section{Advertisers Characterization}
\label{sec:adv}

One of the main goals of this paper is to understand to which extent small and medium-sized businesses (SMEs) are using the Facebook ad platform compared to large-sized businesses (BEs) and analyze their type of activity. 

\subsection{Analysis of businesses by size} 
\begin{figure}
    \centering
    \begin{minipage}{0.49\textwidth}
        \centering
        \includegraphics[width=6.1cm,keepaspectratio]{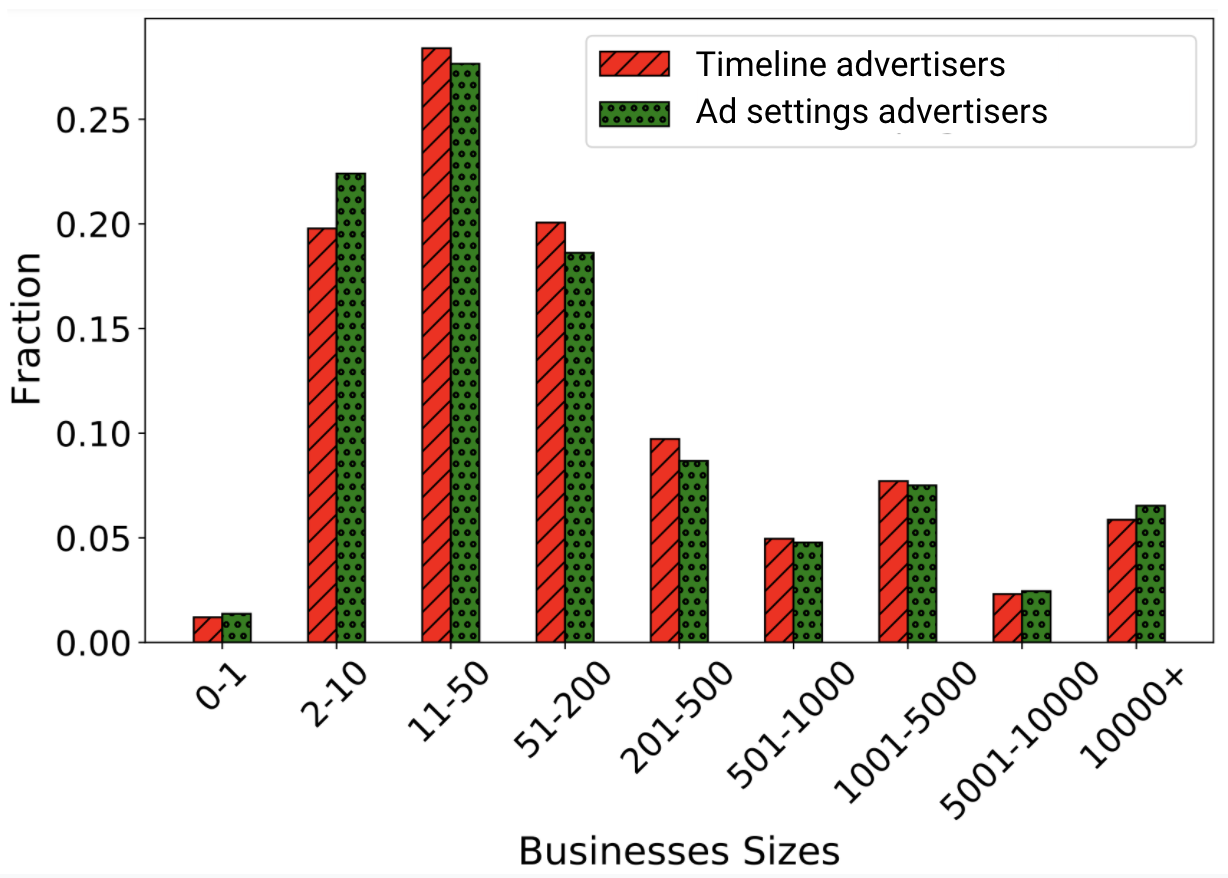} % first figure itself
        \caption{Fraction of advertisers with different business size for \tadv and \padv.}
        \label{fig:sizeb}
    \end{minipage}\hfill
    \begin{minipage}{0.49\textwidth}
        \centering
        \includegraphics[width=6.1cm]{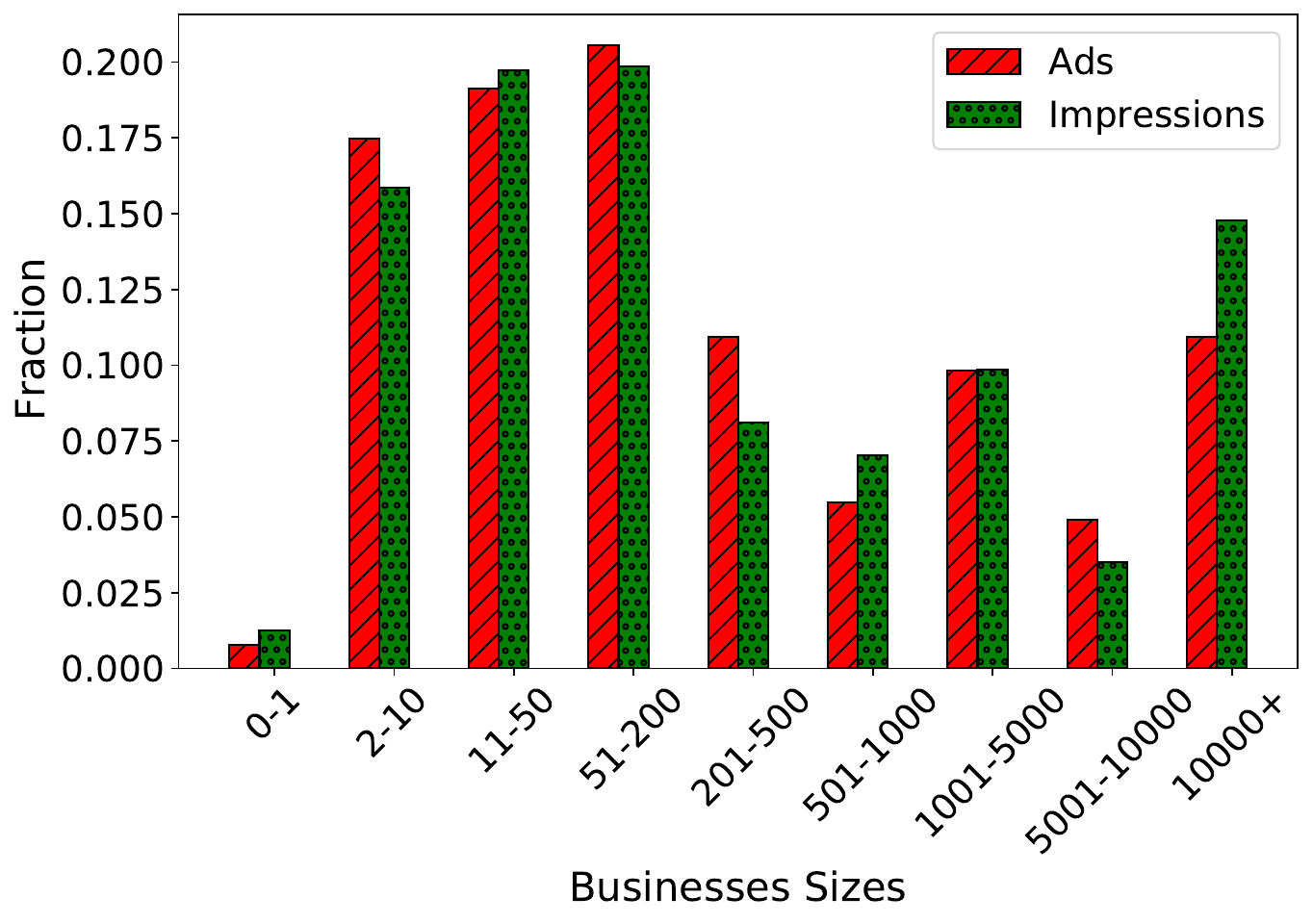} % second figure itself
        \caption{Fraction of ads and fraction of ad impressions by business size for \tadv.}
        \label{fig:adsimp}
    \end{minipage}
\end{figure}

LinkedIn provides information about the number of employees for each matched advertiser. Figure~\ref{fig:sizeb} presents the histogram of the size of businesses that advertise on Facebook for both \tadv and \padv  (for which we have a matching LinkedIn profile). The figure shows that 71\% of \tadv and 74\% \padv  are businesses with 200 employees or less (SMEs). The numbers are consistent across the two advertiser's datasets, which provides confidence the results generalize. Besides, these fractions do not account for the \uadv that are likely small businesses as well. 
If we consider all \tadv, 58\% are SMEs, 24\% are BEs, and 18\% do not have a LinkedIn profile. 
If we consider all \padv, 37\% are SMEs, 13\% are BEs, and 50\% do not have a LinkedIn profile.

While, in numbers, SMEs clearly account for the majority of the advertisers in our dataset, the question is whether SMEs, in aggregate, spend more or less money than BEs on advertising on Facebook. While we do not have information about the amount of money spent on ads by advertisers, we have data about the number of ad impressions (i.e., the number of times the ad appeared on the screen of users in our dataset). If we assume that BEs did not pay more to reach the users that installed our browser extension than SMEs, ad impressions can be considered as a proxy for the ad budget. 

The 32.5k \tadv with a LinkedIn profile, are responsible for 78k ads (out of 102k) and for 130k ad impressions (out of 185k). The 7.5k \tadv with no LinkedIn profile, are responsible for a total of 24k ads (23\% of the total ads in our dataset) and for 55k ad impressions (30\% of the total ad impressions in our dataset).
Figure~\ref{fig:adsimp} presents the fraction of ads and the fraction of ad impressions that come from businesses of different sizes in \tadv (for which we found a match on LinkedIn).  
The figure shows that 61\% of the ads and 57\% of ad impressions come from SMEs while 39\% ads and 43\% ad impressions come from BEs.  Hence, SMEs do seem to account for the majority of ad impressions in our dataset. 

If we consider all \tadv, 40\% of ad impressions come from SMEs, 30\% come from BEs, and 30\% from advertisers without a LinkedIn profile. 
These results support economic theories that predict micro-targeting to be a powerful marketing tool for SMEs. 

In our dataset, the median number of ads sent by SMEs is 1, while for BEs is 3. Hence, BEs send in general more ads than SMEs. This is expected as BEs might have more products and bigger advertisement expenditure. In terms of ad impressions, the median number of impressions for SMEs is 1, while for BEs is 5. In median, \uadv have 1 ad and 1 ad impression which is similar with SMEs.

\subsection{Analysis of businesses by activity}
Each advertiser needs to create a Facebook page and select categories that best describes their activity. We took a random sample of 1000 \madv and we collected the categories mentioned on their Facebook pages.\footnote{The Facebook API limits our requests to 250/day, it would take more than 160 days to collect categories for all advertisers.} 
Table~\ref{tab:industry} shows the most frequent categories mentioned by SMEs and BEs in our dataset. The granularity of the categories mentioned varies across advertisers, with many choosing not to give more precise information than ``Product/Service''.   The table shows that a wide range of industries are advertising on Facebook ranging from ``Education'' to ``Clothing'' and to ``News \& Media''. 

\begin{table}[htp]

\caption{Distribution of advertisers by activity (across both \tadv and \padv).}
\begin{center}
\small{
\begin{tabular}{lcccc}
\toprule
Activity/Industry &   SME (\%) &         BE (\%) & UNMATCHED (\%)\\
\midrule
Product/Service       &  21 &  21   & 24\\
Clothing              &  10 &   7    & 0\\
Beauty \& Cosmetics    &   9 &   9   & 0\\
Retail \& Store        &   7 &  16   & 24 \\
News \& Media          &   6 &   7   & 1    \\
Nonprofit \& Community &   6 &   5   & 10 \\
Education             &   4 &   2    & 0 \\
Entertainment \& Art   &   3 &   4   & 5 \\
Health \& Medics       &   3 &   3   & 1 \\
Home                  &   3 &   1    & 0 \\ 
App                   &   2 &   2   & 2 \\
Other        &   26 &  23   & 33 \\
\bottomrule

\end{tabular}
}
\label{tab:industry}
\end{center}
\end{table}

Table~\ref{tab:industry} presents in 4th column the categories listed by the  \uadv (we took a sample of 500). The top three categories are: ``Product/Service'' (24\%), ``Retail and Store'' (24\%) and ``Non profit and Community'' (10\%). Overall, these categories, differ from SMEs and BEs. 
Since these categories are quite broad, we took a sample of 100 \uadv and check manually their business. 
We observed that 28 corresponded to personal Facebook pages, 24 to services offered by individuals like pet training, housekeeping, 24 to personally owned stores, and 9 to non-profit organizations. These activities seem to represent fringe activities or individual entrepreneurs that are probably too small or too occasional to necessitate the creation of a LinkedIn profile. However, despite their size, they put effort and money in advertising on Facebook.

\vspace{2mm}
\noindent \textbf{Takeaways:}
One of the main benefits cited in the economics literature is that the relative low cost of micro-targeting enables smaller businesses to access advertising markets from which they were previously excluded. This increases the fairness of competition between small and large-sized businesses and lead to a more competitive market. 
Our analysis shows that, indeed small and medium-sized businesses account for the majority of advertisers on Facebook (as captured by our dataset), and they account for the majority of the ad budget spend on Facebook (under the assumption that the price for an ad impression is not higher in average for BEs than SMEs).
Moreover, an increase in the variety of products a user is exposed to, as well as the easiness to find rare and niche products are also important benefits that lead to consumer surplus. Our results show that there is a large diversity of activities promoted on Facebook including a fraction of advertisers who are individual entrepreneurs that promote various fringe activities.

%!TEX root = main.tex

\section{Targeting Characterization}
\label{sec:targeting}

This section analyzes to which extent SMEs and BEs employ micro-targeting strategies when they send their ads on Facebook. 
Our tool collects both the ads and the ad explanations given by Facebook in the ``Why am I seeing this ad?'' option~\cite{noauthor_why_nodate}. These explanations (recall from Section~\ref{sec:data}) contain information about the advertiser-driven micro-targeting (i.e., what are the parameters chosen by the advertiser to target the ad). However, they do not provide information about algorithmic-driven micro-targeting (i.e., what users the ad platform deems ``relevant'' to see the ad). Nevertheless, this data is enough for this section as our focus is on how advertisers use the system.

Out of the different targeting options provided to advertisers, recall from Section~\ref{sec:background}, we consider that advertiser-driven micro-targeting happens when advertisers are creating their audience either using attributes such as demographics and interests or use custom audiences. If advertisers only specify the age, gender, and location of the users they want to reach, we do not consider it as advertiser-driven micro-targeting. Finally, we consider lookalike audiences as a form of explicit algorithmic-based micro-targeting.

Table~\ref{tab:targeting} shows for the 45k ads in our \tadv dataset for which we collected the ad explanation how many have been targeted using: generic targeting only, attribute-based micro-targeting, custom audience-based micro-targeting, and lookalike audiences.  The table shows usage statistics separately for SMEs, BEs, and \uadv. We can see that generic targeting only is used in 72\% of ads sent by SMEs, 69\% of ads sent by BEs and 60\% of ads sent by \uadv. 
The table also shows that 12\% of the ads of SMEs, 15\% of the ads of BEs, and 21\% of ads of \uadv are attribute-based micro-targeted. 
Regarding custom-audience-based micro-targeting, both SMEs, BEs and \uadv  target about 7\% of their ads using custom audiences.  
\textit{Hence, the majority (67\%) of ads for which we have an explanation in our dataset were not advertiser-driven micro-targeted}. This is a significant shift from our previous study in 2019 where we found that 64\% of the ads collected were micro-targeted by advertisers~\cite{andreou_measuring_2019}. This suggests that the way advertisers are employing the Facebook ad platform is evolving. 

If we look at the 33K advertisers for which we have at least one ad explanation, 29\% of SMEs, 30\% of BEs, and 46\% of \uadv micro-targeted at least one of their ads. Hence, most of the advertisers in our dataset do not use advertiser-driven micro-targeting. Nevertheless, the \uadv seem to be the ones that rely the most on advertiser-driven micro-targeting.

\begin{table}
\small
    \centering
    \caption{Usage frequency of different targeting options for ads send by small and medium-sized businesses (SMEs),  large-sized businesses (BEs), and \uadv.}
    \begin{tabular}{lccc}
\toprule
 Targeting type &       SME (\%) &        BE (\%) & Unmatched (\%)\\
\midrule
Generic targeting only &  \textbf{72.3} &  \textbf{69.5} & \textbf{60.6}\\
Attribute-based micro-targeting  &  12.0 &  15.1 & \textbf{21.5}\\
Custom audience-based micro-targeting &  6.9 &  6.8 & 6.6\\
Lookalike audiences &  8.8 &  8.6 &11.2 \\

\bottomrule
\end{tabular}
    \label{tab:targeting}
\end{table}

Targeting based on lookalike audiences is a good example where ad platform's algorithmic-driven micro-targeting interferes \textit{explicitly} in the micro-targeting process. While the advertiser gives the seed audience, it is Facebook's algorithm who selects similar users to create the final lookalike audience. Besides specifying the initial seed audience, advertisers have no control over how this audience is created, and have no idea what are the characteristics of the resulted audience. In our dataset, ~9\% of ads from SMEs and BEs, and 11\% of ads from \uadv are targeted using lookalike audiences.  
Hence, in our dataset, \textit{9\% of ads for which we have the explanation are targeted using explicit algorithmic-driven micro-targeting}.

\paragraph{Zoom on generic targeting}
Generic audiences are defined using location, language, age, and gender. The targeting explanation that Facebook provides with ads allows us to study generic attributes in detail. 81.7\% of ads target all gender groups, while 14.3\% of the ads target women only. As for the age, an advertiser can specify the minimum and maximum age of the targeted audience.  
Half of the ads target users between 18 and 65, which is the default configuration on Facebook ads manager; while the other half have custom configurations of age. In 92\% of the ads, the location was set to the whole country. The 8\% left is distributed over different U.S. states. There is no particular difference between different types of advertisers regarding the use generic parameters.

\paragraph{Zoom on attribute-based micro-targeting}
In the whole period of study, 12k \madv (for which we have at least one ad with an ad explanation) targeted users using 4947 different attributes while the 5k \uadv targeted users with 6219 attributes. SMEs used 3082 distinct attributes to target their ads and BEs used 2949. In median, an advertiser uses three different attributes to target users across different ads. The top five targeted interests for SMEs are online shopping, physical fitness, cats, shopping and fashion and amazon; for BEs are online shopping, culture, shopping and fashion, learning and shopping; and for \uadv are online shopping, shopping and fashion, physical fitness, yoga and cooking. 

\paragraph{Zoom on custom audience-based micro-targeting} An advertiser can form custom audiences from different data sources: a data file containing a list of emails, phone numbers, or users ids (e.g., contact information collected from previous clients), website visits detected by a hidden Facebook pixel (i.e., also traditionally known as re-targeting), users who engaged with the advertiser's page, video, or Instagram profile. Table \ref{tab:cas} shows the different data sources used by advertisers to construct custom audiences. The table shows that data files are more used by \uadv while the Facebook pixel is more used by BEs.

\begin{table}[htp]
\caption{Usage of different data sources for ads targeted using custom audiences.}
\small{
\begin{center}
\begin{tabular}{lccc}
\toprule
Data source &  SME (\%) &  BE (\%) &  Unmatched (\%) \\
\midrule
Data file            &   35 &  33 &         \textbf{54} \\
Website pixel        &   38 &  \textbf{40} &         33 \\
Page engagement      &   17 &   12 &          6 \\
Video engagement     &    7 &   9 &          4 \\
Instagram engagement &    3 &   6 &          3 \\
\bottomrule
\label{tab:cas}
\end{tabular}
\end{center}
}
\end{table}

\vspace{2mm}
\noindent \textbf{Takeaways:} Our results show that only 28\% small and medium-sized businesses, 30\% of large-sized businesses, and 46\% of advertisers without a LinkedIn profile micro-targeted at least one of their ads. In fact, 67\% of the ads in our dataset have been targeted with only age, gender, location, and language parameters. 
However, it is important to understand that these results do not imply that micro-targeting is not useful as a marketing strategy; they rather suggest that the advertisers prefer to outsource the micro-targeting task to the ad platform. Recall from, Section~\ref{sec:background} that Facebook claims to perform algorithmic-driven micro-targeting for every ad impression. The shift away from advertiser-driven micro-targeting could be explained by ad platforms' algorithms getting better than advertisers at finding the right audiences to target ads.
 
Previous work showed that advertisers who target broad audiences might end up ceding platforms even more influence over which users ultimately see the ads~\cite{Ali2021}, which indicates that algorithmic-driven micro-targeting has a more significant role in ads targeted at large audiences. Our work shows that most ads on Facebook use large audiences, pointing to algorithmic-driven micro-targeting as a significant driver of delivering ads.

\section{The role of web tracking in micro-targeting on Facebook}
\label{sec:tracking}

Civil societies are pushing lawmakers to ban web tracking that is intended for advertising~\cite{edri} and big platforms such as Apple and Google are also taking technical steps to make it harder to track users online. While the role of web tracking in online advertising is widely acknowledged~\cite{6234427}, the role of web tracking in advertising on Facebook is less clearly understood as the platform already has a lot of information about the behavior of users on their social network.

\subsection{Mentions of web tracking in Facebook's policies and help pages}
To better understand the role of web tracking  in micro-targeting on Facebook we first turned to the policies and help-pages provided by Facebook:

For \emph{attribute-based advertiser-driven micro-targeting}, Facebook states in their ``Detailed Ad Targeting'' help page that the demographics, interests and behaviors made available to advertisers to select their audiences are based on activity \emph{inside} Facebook such as their comments and the interactions with different contents, and do not mention any outside signals coming from web tracking~\cite{facebook_business_about_detailed_targeting}: 
\begin{displayquote}
\small{\em ``Ads they click;
Pages they engage with;
Activities people engage in across Meta technologies related to things like their device usage, and travel preferences;
Demographics like age, gender and location;
The mobile device they use and the speed of their network connection.''}
\end{displayquote}

Hence, attribute-based advertiser-driven micro-targeting \emph{a priori}, does not rely on web tracking.   

When looking at \emph{custom audiences-based advertiser-driven micro-targeting}, the only option that employs web tracking is the option of targeting users that visited the advertiser's website, i.e., retargeting. This is done through the Facebook pixel, a piece of code that advertisers can install on their website and that allows Facebook to track what users are visiting on the advertiser's website. 

We could not find any information regarding the use of web tracking in the \emph{lookalike audience}.

Regarding \emph{algorithmic-driven micro-targeting} we found two mentions of the use of web tracking. First mention is in the ``Ad principles'' page that clearly states that Facebook uses information about what websites users visits to choose to whom to send an ad ~\cite{facebook_business_ad_principles}:
\begin{displayquote}
\small{\em
``Our system can show the hair salon's ads to the people in that area who are most likely to be interested in the salon. Interests are determined based on activity on Facebook, such as Pages or posts that people like, posts or comments they make and activity off Facebook from apps they use and \emph{websites they visit}.'' }
\end{displayquote}

The second mention is in the ``Facebook Business Tools Terms'' (i.e., the terms a business needs to sign to use the Facebook pixel or other social plugin such as the Like button) that mention they use information gathered from businesses through Event Data to determine the relevance of ads to people~\cite{facebook_business_tools_terms}:

\begin{displayquote}
\small{\em
``We may correlate that Event Data to people who use Facebook Company Products to support the objectives of your ad campaign, improve the effectiveness of ad delivery models, and determine the relevance of ads to people. We may use Event Data to personalize the features and content (including ads and recommendations) that we show people on and off our Facebook Company Products.''}
\end{displayquote} 
Hence, it seems that web tracking does not play a big role in advertiser-driven micro-targeting, but web tracking is used in algorithmic-driven micro-targeting. Since Facebook claims to perform algorithmic-driven micro-targeting for every ad delivery, web tracking seems to be an important source of data for micro-targeting every ad on Facebook~\cite{facebook_business_about_ad_delivery}.

\subsection{Prevalence of the Facebook pixel among advertisers}
Web tracking data can come either from websites that are implementing Facebook's tools such as the Like button, either from advertisers that install a Facebook pixel. 
To understand the extent to which advertisers contribute to Facebook's data acquisition through web tracking, we investigate what fraction of advertisers installed the Facebook pixel on their website.
Facebook pixel is a tracker that allows advertisers to track and analyze visitors' activity on their websites~\cite{noauthor_facebook_pixel}. 
When the user visits the advertiser's website, the pixel sends the event of ``page view'' to Facebook. If the user clicks on the purchase or the buy button, the pixel generates a purchase event~\cite{noauthor_conversion_events}. All the data about different events on the website (visits, purchases, subscriptions) are sent to Facebook. The advertisers have access to a report about those events on the Facebook Ads Manager. Advertisers can use the data gathered from their installed Facebook pixels to target the visitors of their website.

To estimate the fraction of businesses that have deployed a pixel on their websites we took a random sample of 600 advertisers (61\% SMEs, 39\% BEs) from our database, and we checked their websites manually for potentially installed pixels. To see if a website uses a Facebook pixel or not, Facebook provides a browser extension called Facebook Pixel Helper~\cite{facebook_pixel_helper}. We only took a limited sample of 600 advertisers because we cannot fully automate checking the pixel. When visiting a website for the first time, a user must accept cookies and consent to the terms of the website's privacy policy, including accepting third-party trackers like Facebook Pixel. Accepting cookies cannot be automated easily since each website implements consent forms (cookies configuration) differently. Thus, we were obliged to do it manually and hence the limitation.

Our analysis shows that 81\% of SMEs and 69\% of BEs have deployed one or more pixels on their websites (19\% have more than one pixel). 91\% of advertisers only track the website visits (page view) while 9\% configure their pixels to track more events like purchase.  

\vspace{2mm}
\noindent \textbf{Takeaways:} Facebook mentions the use of web tracking in algorithmic-driven micro-targeting but not in attribute-based advertiser-driven micro-targeting. This could factor in why advertisers seem to prefer algorithmic-driven micro-targeting over advertiser-driven micro-targeting.  
Furthermore, 81\% of the SMEs and 69\% of the BEs we checked in our dataset are sending data about their website visitors to Facebook. Nevertheless, only 2.6\% of the ads sent by SMEs and 2.7\% of the ones sent by BEs were targeted explicitly with the Facebook pixel to reach people that visited the advertiser's website. This raises questions of whether advertisers have any benefits from installing the Facebook pixel, besides targeting people who visited their website.

%!TEX root = main.tex

\section{Related Works}

\subsection{Economics of online advertising}
The economics literature on online advertising can be split in two categories: 
(1) studies on modeling online advertising and its benefits and consequences for users and firms, and (2) studies about the risks brought by highly concentrated marketing structures with only a few players that offer online advertising services.

From a conceptual perspective, advertising can be characterized as $i$) persuasive, altering consumer tastes; $ii$) informative, reducing the cost of information acquisition by consumers; $iii$) complementary to the advertised product, increasing the consumption value of a product without altering underlying preferences~\cite{bagwell2007economic}. Online advertising fits into this definition, and many researchers advocate that the fundamental difference compared to offline advertising is the development of micro-targeting that leads to a substantial reduction in the cost.

Many theoretical models have portrayed the potentially positive economic consequences of micro-targeting development: 

\setlist{nolistsep}
\begin{enumerate}
  \item \textit{Increasing the demand}: By improving the content and efficiency of messages, micro-targeting should lead to an \textit{increase in demand} as consumers are ready to pay more for a product that perfectly fits their preferences~\cite{athey2010impact}.
  
 \item \textit{Increasing the market competition}: The micro-targeting expansion and, more specifically, the consecutive \textit{reduction in consumer search costs} should result in consumers being better informed, which allows them to compare prices and products more quickly and contributes to the competition between advertisers.
    
  \item \textit{Decreasing the prices of products}: The relatively low cost of micro-targeting enables smaller businesses to \textit{access advertising markets} from which they were previously excluded~\cite{anderson2006long}. This can help in decreasing the prices as it increases the supply and thus market competition.

  \item \textit{Finding unknown products}: The decrease in consumer search costs also means that it is easier to \textit{find relatively unknown, rare, and niche products}~\cite{yang2013targeted,Zhang2017}. 
\end{enumerate}
   \setlist{listsep}
   \setlist{}

These models rely, however, on simplified assumptions. They assume zero search costs for consumers, whereas empirical results suggest otherwise. Structural estimates of the cost of an extra click in the consumer search process suggest they are more significant than assumed~\cite{honka2014quantifying, De2019}. This means that consumers stop searching sooner than predicted by models that assume search costs close to zero. Thus micro-targeting may have a smaller beneficial impact than the theoretical models suggest as it does not lead to a perfectly informed consumer. 
Moreover, the consumers are facing a trade-off between stopping searching and making up their minds with the available information (\textit{knowing that there is maybe a better product if they keep looking or continue to search}) and endure an effort in continually searching for the perfect fit. 
Finally, privacy-focused consumers react negatively to advertisements that are both targeted and obtrusive ~\cite{goldfarb2011advertising}. Laboratory studies and surveys have even documented consumer discomfort with targeted advertising~\cite{malheiros2012too, tsai2011effect, turow2009americans}.
%Few empirical findings have tried to test those different predictions. \oana{what predictions?} However,  \oana{this is consistent with theoretical models no?}

Finally, the source of targeting, namely large amounts of information about consumers, may often lead to highly concentrated market structures, as is the case in search and social networks such as with Google and Facebook~\cite{bergemann2011targeting}. Overall, ad platforms are becoming subject to more antitrust scrutiny~\cite{goldfarb2011advertising,goldfarb2011online}. The arguments for this regulatory attention rely on privacy data matters and on the targeting techniques used by the search and social networks. While online advertisements and micro-targeting are especially valuable to smaller advertisers that do not have easy access to offline advertising, the configuration of the algorithms employed by ad platforms to deliver ads may substantially impact smaller advertisers. For example, \citet{brynjolfsson2003consumer} showed that online advertising might alter the fairness of competition between advertisers as it can produce different consumer search costs according to the recommendation algorithm or the micro-targeting strategies used.
Furthermore, using data from eBay, ~\citet{dinerstein2018consumer} emphasizes how the design of the search algorithm on eBay affects markups charged by eBay sellers. Hence, micro-targeting should be accessible for all advertisers in the same way to ensure fair competition. 
Thus, on top of the micro-targeting risk to be weaponized for political purposes, there is a clear call for more antitrust scrutiny regarding the fairness of use and access of micro-targeting by advertisers, especially small and medium-sized businesses.

\subsection{Empirical studies on online advertising}

Many works have been published in recent years about the Facebook advertising platform. These works can be categorized into: (1) studies about the Facebook advertising ecosystem in general~\cite{andreou_measuring_2019, dehghani2015research, brettel2015drives, hutter2013impact, erkan2019impacts}, (2) studies about the targeting of political ads on Facebook~\cite{Ghosh2019AnalyzingPA, Ali2021, silva2020facebook, Sosnovik_2021, le_ponchat_277076}, and (3) studies about the risks brought by online advertising~\cite{Ali_2019, speicher_potential_2018, Caba_as_2021}.

The closest to this work is our previous study (Andreou et al.~\cite{andreou_measuring_2019}), where we analyzed the data collected from 600 users who installed the AdAnalyst browser extension~\footnote{\url{https://adanalyst.mpi-sws.org}}. Our previous results describe the usage of Facebook targeting options by different advertisers and provide a detailed statistical summary of the targeted population, targeting options, and the activity of advertisers. Some works conducted studies on the effectiveness of Facebook advertising platform on increasing purchases~\cite{dehghani2015research, brettel2015drives} and brand awareness~\cite{hutter2013impact, erkan2019impacts}. 

Many recent works focused specifically on political ads: Ghosh et al. \cite{Ghosh2019AnalyzingPA} worked on the targeting of such ads and found that well-funded advertisers tend to use privacy-sensitive targeting features more frequently while less-well-funded advertisers tend to more narrowly target their audiences geographically. Also, Ali et al. \cite{Ali2021} investigated the impact of Facebook's ad delivery algorithms on political ads. They found that these algorithms differentiate the price of reaching a user based on their inferred political alignment with the advertised content, inhibiting political campaigns' ability to reach voters with diverse political views, which leads to political content polarization. Silva et al. \cite{silva2020facebook} studied the detection of political ads; after monitoring the political ads during the 2018 Brazilian elections, they noted that a significant fraction of political ads they detected were not present in the Facebook political ad library, which means that they were not declared as political. This emphasizes the importance of enforcement mechanisms for declaring political ads and the need for independent auditing platforms. Sosnovik et al.~\cite{Sosnovik_2021} showed that there is a significant disagreement between ad platforms, ordinary people, and advertisers on what is political. This indicates how difficult it is to detect political ads and distinguish them from non-political ones reliably. Finally, Le Pochat et al.~\cite{le_ponchat_277076} studied the enforcement decisions made by Facebook for ads that may be political but were not declared as such by the advertiser. They found that the current enforcement is imprecise (61\% more ads are missed than are detected) and the detection performance is uneven across countries.

%There is also a website where we can check some of the ads people disagree on\footnote{\url{https://facebookads.imag.fr/}}  

Finally, more broadly, several works have looked at other risks with online advertising platforms. Ribeiro et al. \cite{Ribeiro_2019} studied how the Russian Intelligence Research Agency (IRA) has interfered in the 2016 US presidential elections. They showed that the very efficient ad targeting done by the Russian Intelligence Research Agency (IRA) was to a good extent enabled by the enormous amount of personal data aggregated by Facebook on users and made available for advertisers. Another risk of using online advertising is discrimination by targeting only specific groups with interesting opportunities. Ali et al. \cite{Ali_2019} show that the market and financial optimization of Facebook's algorithms can lead to skewed delivery outcomes. Also, Speicher and al. \cite{speicher_potential_2018} investigate in their work the different targeting methods offered by Facebook and showed that some of them could be used by a malicious advertiser to create highly discriminatory ads without specifying any sensitive attributes. This proves that the measures taken by Facebook - disallowing the use of attributes such as ethnic affinity from being used by advertisers - are not sufficient to stop targeting discrimination. To limit these risks, some data protection regulations restrict the processing of some categories of personal data (ethnic origin, religious beliefs, etc.) due to the privacy risks associated to such information. However, Cuevas et al. \cite{Caba_as_2021} show that this enforcement had a negligible impact by showing that the users labeled with such personal interest remained the same few months after the GDPR was enacted.

While other works focus on separate sides of the platforms and sometimes limited to an actor or a process, our study makes the link between all those in a data analysis setup. Using data, we were able to give a clear idea of the nature and the activity of advertisers on the platform and their targeting practices. This is a first step towards understanding the benefits of online advertising platforms from an economical perspective.

%!TEX root = main.tex

\section{Concluding Discussion}
In this work, we conducted a study about advertisers on Facebook and their targeting practices. Combining the data collected by our browser extension and the data from LinkedIn, we could profile advertisers based on their company size and targeting activity on Facebook. Our study included 40K advertisers who targeted 890 U.S. users. 
We discovered a strong presence of small and medium-sized businesses (SMEs) on Facebook, accounting for 71\% of advertisers in our dataset. SMEs are also responsible for 57\% of ad impressions. However, neither SMEs nor B.E.s rely heavily on micro-targeting; 67\% of ads in our dataset were targeted by advertisers using only age, gender, location, and language attributes. 

Facebook is optimizing ad delivery for every ad on top of the targeting specified by the advertisers. One possible explanation for why advertisers choose not to micro-target their ads to specific groups of users is that Facebook's algorithms are doing a better job at finding the right users to show the ads without the input of advertisers. While most discussions until now have focused on micro-targeting done by advertisers, our work shows that it is equally important to give attention to micro-targeting done by ad platforms. The current lack of transparency of algorithmic-driven micro-targeting leads to an inability to assess the fairness and safety of this type of micro-targeting for both users and advertisers.

Our paper also contributes to a better understanding of the role of web tracking in Facebook's micro-targeting options.
While web tracking does not seem to contribute a lot to advertiser-driven micro-targeting, it does seem to play a role in algorithmic-driven micro-targeting.
Our analysis showed that the majority of the advertisers we checked, irrespective of their size, are allowing Facebook to track users that visit their websites by installing a Facebook pixel. 
This pixel is an excellent source of data for Facebook as it gives information about the behavior of users outside Facebook and provides precious information about what users are interested in at the current moment.

Our work does not confirm the usefulness or non-usefulness of micro-targeting for advertisers. However, it suggests that on Facebook, the bulk of the micro-targeting might be done by the ad platform and not by the advertisers themselves. It is important to take this knowledge into account in future legislation and not focus solely on micro-targeting done by advertisers. From an economics perspective, our work also opens questions on whether algorithmic-driven micro-targeting leads to a competitive market (e.g., is it exposing consumers to more options and allowing them to make more environmental-friendly choices), and what are the new risks for both users and advertisers.

\section*{Acknowledgements}
This research was supported in part by the French National Research Agency (ANR) through the ANR-17-CE23-0014 and the MIAI@Grenoble Alpes ANR-19-P3IA-0003 grants and the European Union’s Horizon 2020 research and innovation program under grant agreements No 101021377 and No 952215. 

%Partially supported by TAILOR, a project funded by EU Horizon 2020
%research and innovation programme under GA No 952215

%%
%% The acknowledgments section is defined using the "acks" environment
%% (and NOT an unnumbered section). This ensures the proper
%% identification of the section in the article metadata, and the
%% consistent spelling of the heading.
%\begin{acks}
%To Robert, for the bagels and explaining CMYK and color spaces.
%\end{acks}

%%
%% The next two lines define the bibliography style to be used, and
%% the bibliography file.
\bibliographystyle{ACM-Reference-Format}
\bibliography{bib}

% For CSCW2 Article 356-378, use
\received{July 2021}
\received[revised]{November 2021}
\received[accepted]{February 2022}

\appendix

\newpage
\section{Appendix}

\subsection{Ad explanations}

\begin{figure}[h]
     \centering
     \begin{subfigure}[b]{0.3\textwidth}
         \centering
         \includegraphics[width=\textwidth]{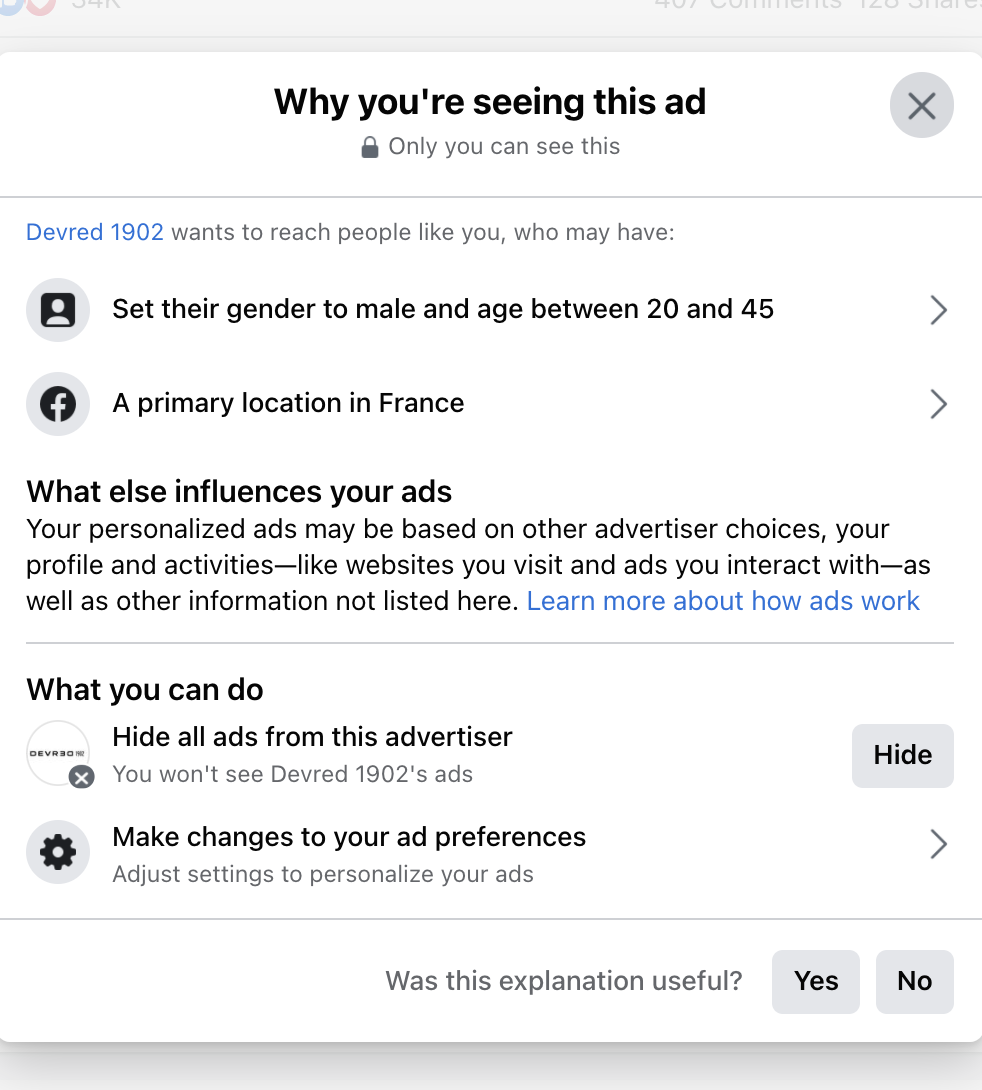}
         \caption*{Generic targeting explanation (only age, gender and location)}
     \end{subfigure}
     \hfill
     \begin{subfigure}[b]{0.3\textwidth}
         \centering
         \includegraphics[width=\textwidth]{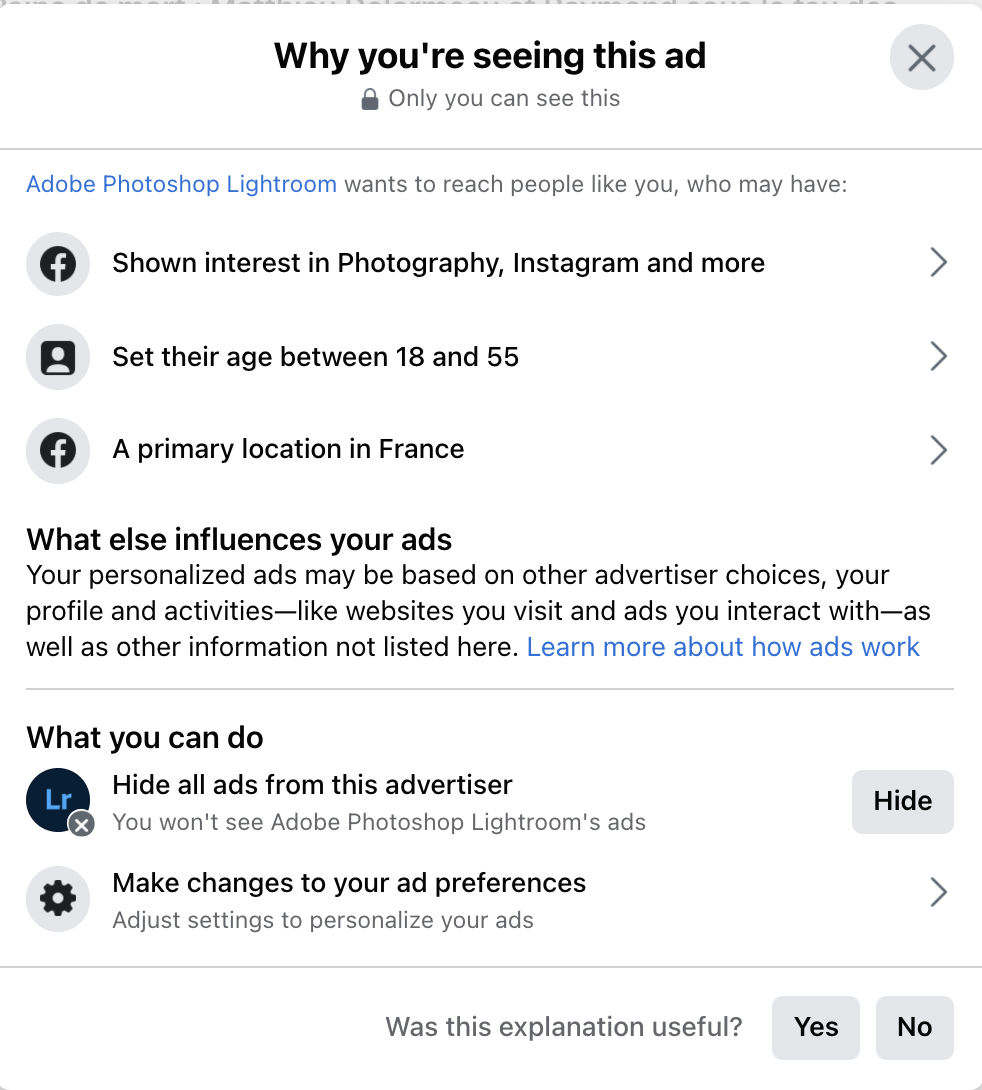}
         \caption*{Attribute-based targeting explanation}
     \end{subfigure}
     \hfill
     \begin{subfigure}[b]{0.3\textwidth}
         \centering
         \includegraphics[width=\textwidth]{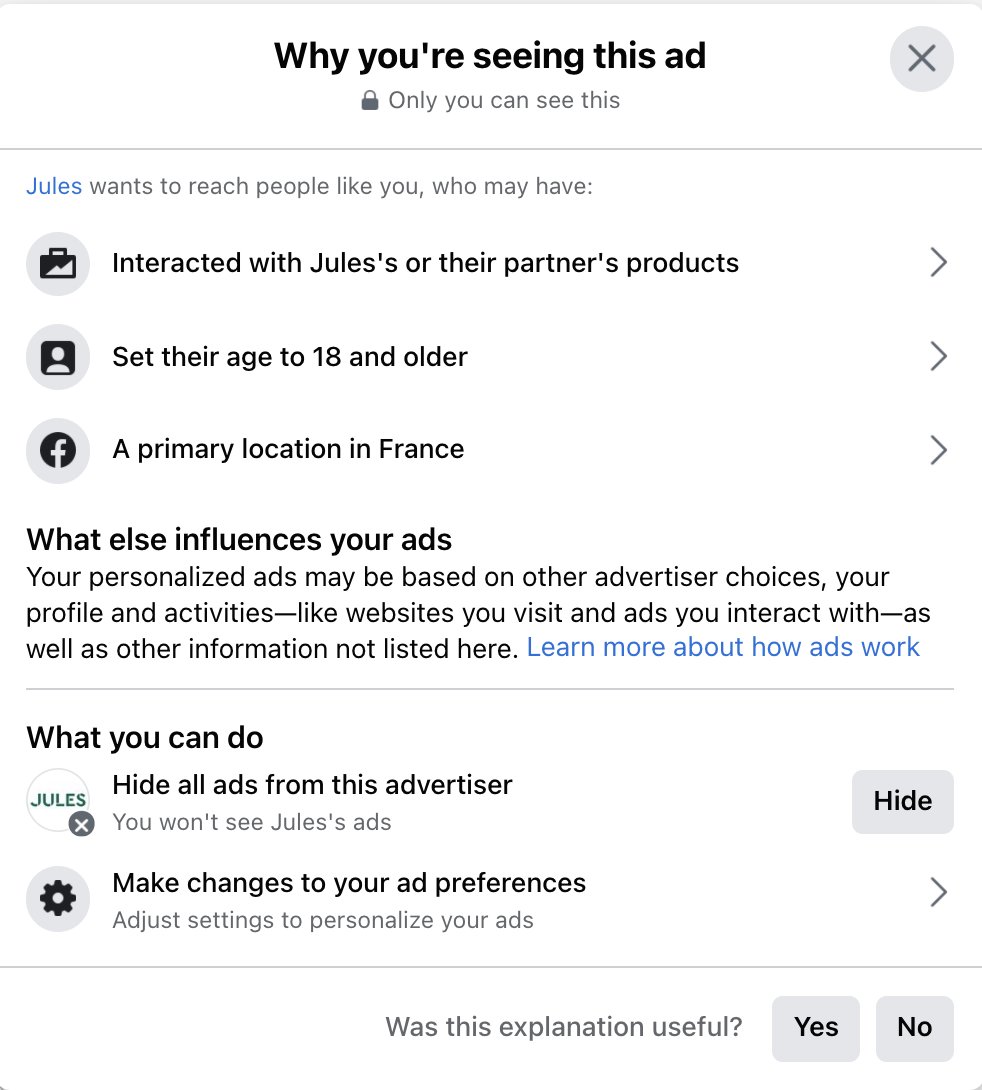}
         \caption*{Custom-audience pixel-based targeting explanation}
     \end{subfigure}
     
    \caption{Screenshot of different ads explanations as shown to users.}
    \label{fig:explanation}
\end{figure}

\subsection{Representativeness}

\begin{figure}[h]
    \centering
    \begin{minipage}{0.49\textwidth}
        \centering
        \includegraphics[width=6.5cm,keepaspectratio]{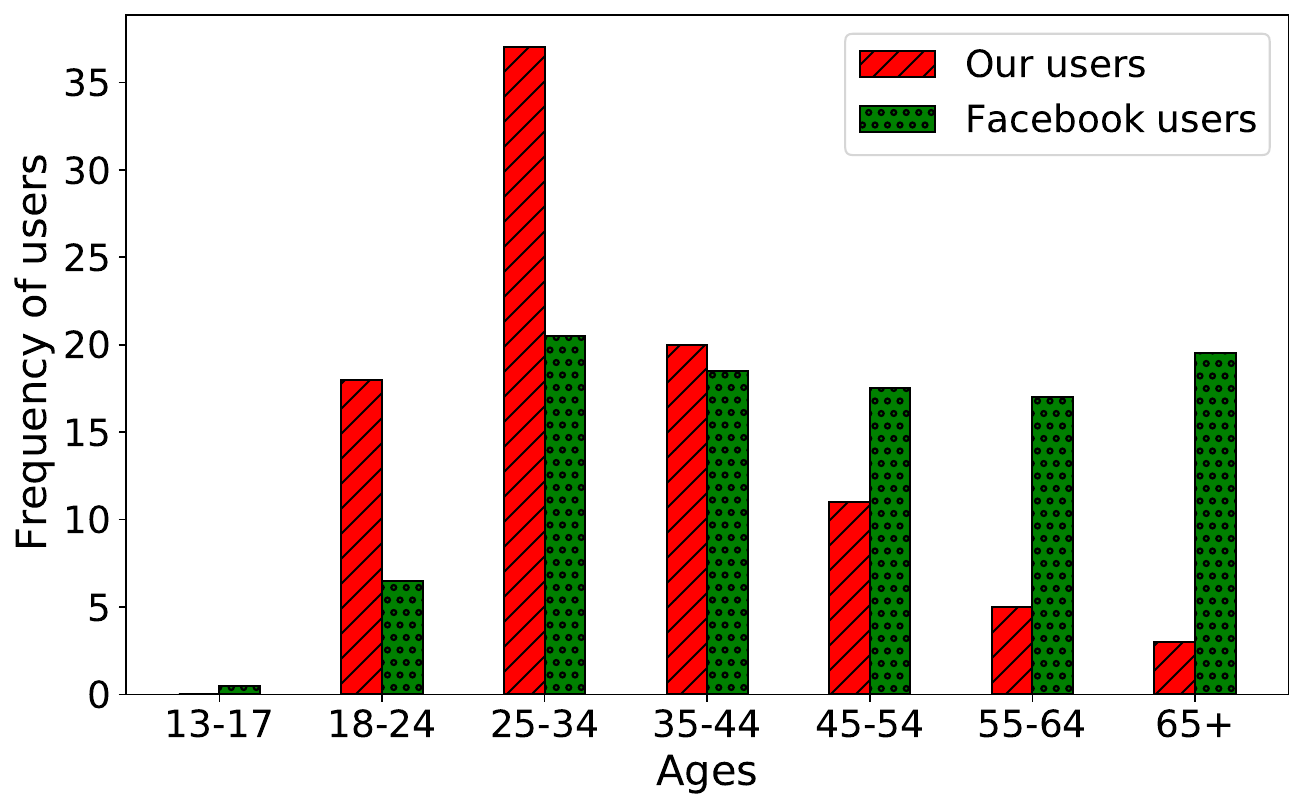} % first figure itself
        \caption{Comparison of users ages between our set of users and Facebook users.}
        \label{fig: agesdist}
    \end{minipage}\hfill
    \begin{minipage}{0.49\textwidth}
        \centering
        \includegraphics[width=6.4cm,keepaspectratio]{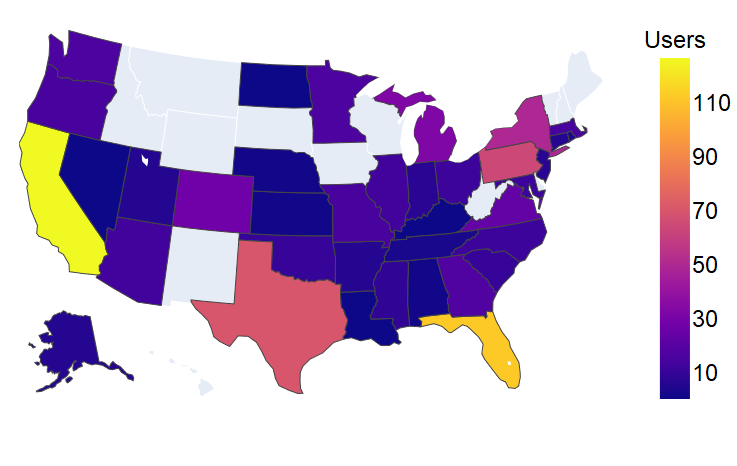} % second figure itself
        \caption{Location of our users.}
        \label{fig: usmap}
    \end{minipage}
\end{figure}

%%
%% If your work has an appendix, this is the place to put it.
%\appendix
%\section{Research Methods}

\end{document}